\documentclass [conference]{IEEEtran}
\IEEEoverridecommandlockouts
\usepackage{cite}
\usepackage{amsmath,amssymb,amsfonts}
\usepackage{comment}
\usepackage{algorithmic}
\usepackage{graphicx}
\usepackage{textcomp}
\usepackage{xcolor}
\usepackage{graphicx}
\usepackage{float}
\usepackage{array}
\usepackage[utf8]{inputenc}
\usepackage[T1]{fontenc}

\DeclareUnicodeCharacter{221A}{\ensuremath{\sqrt{}}}   
\DeclareUnicodeCharacter{2202}{\ensuremath{\partial}} 

\def\BibTeX{{\rm B\kern-.05em{\sc i\kern-.025em b}\kern-.08em
    T\kern-.1667em\lower.7ex\hbox{E}\kern-.125emX}}
    
\begin{document}

\title{Imaging Modalities-Based Classification for Lung Cancer Detection\\
}

\author{\IEEEauthorblockN{Sajim Ahmed$^{1}$, Muhammad Zain Chaudhary$^{1}$, Muhammad Zohaib Chaudhary$^{1}$, Mahmoud Abbass$^{1}$, \\ Ahmed Sherif$^{1}$ (IEEE, Senior Member), Mohammad Mahbubur Rahman Khan Mamun$^{2}$}
	
 \IEEEauthorblockA {$^{1}$ School of Computing Sciences and Computer Engineering,  University of Southern Mississippi, MS, USA
		\\
        $^{2}$ Electrical and Computer Engineering Department, Tennessee Technological University, TN, USA
       } 
       \\
		Emails: \{sajim.ahmed@usm.edu, zain.chaudhary@usm.edu, zohaib.ali@usm.edu, mahmoud.abbass@usm.edu, \\   ahmed.sherif@usm.edu, mmahmoud@tntech.edu\}}


\maketitle

\begin{abstract}
Lung cancer continues to be the predominant cause of cancer-related mortality globally. 
This review analyzes various approaches, including advanced image processing methods, focusing on their efficacy in interpreting CT scans, chest radiographs, and biological markers. 
Notably, we identify critical gaps in the previous surveys, including the need for robust models that can generalize across diverse populations and imaging modalities. 
This comprehensive synthesis aims to serve as a foundational resource for researchers and clinicians, guiding future efforts toward more accurate and efficient lung cancer detection.
Key findings reveal that 3D CNN architectures integrated with CT scans achieve the most superior performances, yet challenges such as high false positives, dataset variability, and computational complexity persist across modalities.

\end{abstract}


\section{Introduction}

Lung cancer remains one of the leading causes of cancer-related deaths worldwide, with nearly 1.8 million deaths annually \cite{WHO2020}. 
Early detection and accurate diagnosis are vital for improving lung cancer survival, as early-stage cases are more treatable. 
While traditional methods like chest X-rays and CT scans have been key in detection, they face challenges in sensitivity and accuracy for early-stage malignancies. 
Recent advances in medical imaging and deep learning techniques show great promise in addressing these limitations \cite{Z10, Z11, Z12}.
In recent years, computer-aided detection (CAD) systems have played a crucial role in enhancing the accuracy and speed of lung cancer diagnosis. 
When integrated with traditional imaging modalities such as CT scans and PET/CT, these systems significantly improve detecting and classifying lung nodules \cite{Z4, S5, S9}. 
Several studies have successfully applied CNN architectures, such as Inception v3 and U-Net, to lung cancer detection tasks \cite{S3, Zo9}.

Despite the advancements in imaging techniques, several challenges remain in lung cancer detection. 
False positives, computational complexity, and dataset limitations are some of the significant obstacles faced by existing models \cite{Zo9, S10}.  
Solutions such as data augmentation, semi-supervised learning, and transfer learning have been proposed to mitigate these challenges. 
However, further research is necessary to refine these methods and improve their practical applications.

This survey aims to comprehensively review the state-of-the-art imaging modalities used in lung cancer detection. 
Unlike prior literature, our work introduces a unified 
By classifying existing research of imaging modalities architectures-based lung cancer detection into three main categories. 
This paper presents an in-depth analysis of the strengths and limitations of each approach. 
Furthermore, we identify open research problems and propose solutions to address these challenges, offering a roadmap for future work in lung cancer detection.

This paper is organized as follows.
Section \ref{relted} discusses related work in the field. 
Section \ref{method} presents the classification method of the reviewed papers, while Section \ref{discussion} offers a detailed discussion of the findings.
Section \label{conclusion} concludes with findings and future research directions.

\begin{figure}
\centering
    \includegraphics [width=0.75\columnwidth]{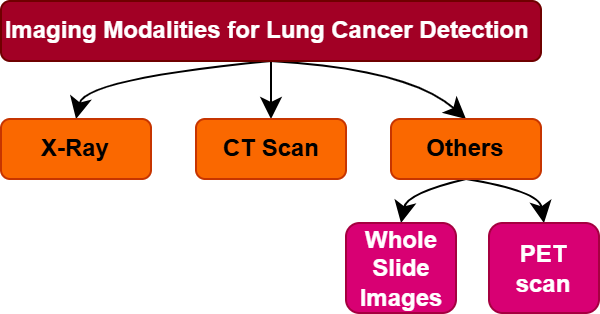}
\caption{Taxonomy of Lung Cancer Detection Techniques.}
\label{fig:taxonomy}
\end{figure}

\section{Related work}
\label{relted}

Several schemes have been proposed for using the emerging technology in E-health \cite{nn1, nn2, nn3, nn4, nn5}.
The identification of lung cancer has been transformed in recent years by using artificial intelligence (AI) in medical imaging. 
The paper \cite{Z10} provides a comprehensive review of AI's role in lung cancer detection, segmentation, and characterization. 
The classification techniques used in this paper center on CNNs and 3D CNNs. 
However, the paper highlights challenges, such as the need for large datasets and persistent false-positive results, even in advanced deep learning (DL) models.
Furthermore, the paper \cite{Z11} reviews the current applications of AI in lung cancer screening. 
Regarding classification techniques, the paper highlights the use of CNNs and U-Net architectures for tasks like nodule detection and segmentation.
Also, The paper \cite{Z12} provides a comprehensive review of the significance of using DL models like CNNs, 3D CNNs, and other hybrid models in analyzing medical data from CT scans, chest X-rays, and other modalities for identifying lung cancer at an early stage.

Our proposed classification scheme offers a more comprehensive and structured approach than other papers by integrating many techniques under the unified framework of imaging modalities. 
Unlike prior literature, our work introduces a unified classification framework that systematically evaluates and contrasts the technical efficacy, clinical applicability, and limitations of different imaging modalities.

\section{Proposed Classification Method}
\label{method}

As shown in Fig. \ref{fig:taxonomy}, the proposed classification scheme in this paper is demonstrated based on imaging modalities as follows.

\subsection{X-ray Scan}

X-ray-based lung cancer detection methods vary significantly in preprocessing strategies, model architectures, and clinical applicability. 
While \cite{S4} achieves 92\% validation accuracy through bone shadow exclusion (BSE) and lung segmentation, \cite{li2020exploring} highlights absorption-contrast imaging as superior for early detection, bypassing algorithmic preprocessing. 
Conversely, \cite{S16} and \cite{S14} rely on PCA and basic segmentation, achieving high training accuracy but lower generalization (96\% and 93.33\%, respectively), underscoring the critical role of advanced preprocessing in reducing false negatives.

Simpler models like the PNN in \cite{S14} (93.33\% accuracy) and MANN in \cite{S15} (72.96\% accuracy) prioritize computational efficiency but suffer from small datasets and sensitivity to noise. 
In contrast, complex architectures like CDC-Net \cite{S13} (99.39\% accuracy) and VGG19-CNN \cite{Zo1} (98.05\% accuracy) leverage large datasets (30,387+ images) and residual networks to enhance robustness, albeit at higher computational costs. 
Hybrid models strike a balance. 
OCNN-SVM \cite{S18} combines CNNs with SVMs for 98.7\% accuracy, while VDSNet \cite{S17} integrates spatial transformers to handle rotated images, achieving 73\% accuracy with reduced training time.
Studies like \cite{Zo1} and \cite{S13} address multi-disease detection (COVID-19, pneumonia, etc.), achieving near-perfect AUC (99.66\%–99.93\%) but risking misclassification from overlapping features. In contrast, \cite{S4}, \cite{S16}, and \cite{S18} focus solely on lung cancer, optimizing precision (e.g., 98.76\% F1-score in \cite{S18}) at the cost of diagnostic scope. 
Hierarchical approaches like \cite{S12} show high initial sensitivity (99\%) but degrade in later stages (78\% accuracy), reflecting a precision-recall trade-off. 
Meanwhile, \cite{S15}’s MANN reduces false positives but suffers from low sensitivity (72.85\%), while \cite{S17}’s VDSNet prioritizes rotation invariance over peak accuracy.

Preprocessing and hybrid models (e.g., \cite{S4}, \cite{S18}) enhance specificity, while large-scale CNNs (e.g., \cite{S13}) improve generalizability at higher computational costs. 
Table I synthesizes these limitations, emphasizing the need for standardized benchmarks and explainable AI to bridge clinical adoption gaps. 

\begin{table}

    \caption{X-Ray}
    \label{tab:my_label1}

    \centering
    \setlength{\tabcolsep}{1mm}
    \newcolumntype{M}[1]{>{\centering\arraybackslash}m{#1}}
    \begin{tabular}{|M{0.5cm}|M{2.75cm}|M{4.75cm}|} \hline 
 Ref. & Methodology & Disadvantages\\ \hline 
         \cite{S4}& Lung segmentation and bone shadow exclusion (BSE).
&  Sensitivity Issues: Poor detection of small or early-stage cancers.
False Negatives: Common due to overlapping structures.\\ \hline 
         \cite{Zo1}&  Multi-classification model for COVID-19, pneumonia, and lung cancer. &   Sensitivity Issues: Overlapping between disease symptoms can cause misclassification.
         False Positives: Possible due to overlapping features in X-rays and CT images.\\ \hline
        \cite{li2020exploring}& Adsoption-contrast and phase-contrast imaging & Adsorption contrast imaging affect patients with higher radiation doses. Phase-contrast imaging is complex and expensive to be implemented.\\ \hline
        \cite{S12} & A CAD algorithm to identify pulmonary nodules.  & 
        Limited to retrospective data, which may not fully represent current clinical scenarios.
        The algorithm's performance heavily depends on the quality of the input X-rays.\\ \hline
        \cite{S13} &  CDC\_Net model that  incorporates residual networks and dilated convolution. & 
        High computational cost due to the complexity of the model.
        Potential overfitting due to the use of multiple pre-trained models.\\ \hline
        \cite{S14} & A probabilistic neural network. & 
        Sensitive to noise in the data.
        The model may not generalize well to different datasets.\\ \hline
        \cite{S15} & CAD system using a massive artificial neural network (MANN) with a soft tissue technique. &
        The model's sensitivity to subtle nodules is still relatively low.
        High false positive rate, which can lead to unnecessary follow-up tests.\\ \hline
        \cite{S16} & Combined backpropagation neural networks with PCA. & 
        PCA may lead to loss of important information during dimensionality reduction.\\ \hline
        \cite{S17} & A hybrid deep learning model combining VGG, data augmentation, and spatial transformer networks with CNN. & 
        The model's performance can degrade with rotated or tilted images.
        Requires significant computational resources for training.\\ \hline
        \cite{S18} & A hybrid deep learning technique combining CNN and SVM. &
        Complex to implement and optimize.
        Large amounts of labeled data are required for effective training. \\ \hline

    \end{tabular}

\end{table}

\subsection{CT Scan}

CT-based lung cancer detection methods exhibit diverse architectural strategies, preprocessing pipelines, and clinical trade-offs. 
Pure 3D CNNs, such as those in \cite{Z4} (U-Net segmentation) and \cite{S5} (dual 3D CNNs for detection/classification), achieve high sensitivity (94\%) and specificity (91\%) by leveraging spatial context, but face computational bottlenecks. 
Hybrid models address this. 
In \cite{Z6}, the paper combines improved deep neural networks (IDNN) with an ensemble classifier, achieving robust accuracy through feature optimization, while \cite{Z9} integrates CNNs with mRMR feature selection to balance efficiency and performance. 
In contrast, \cite{Zo7} reduces computational load via template matching and simplified preprocessing without sacrificing accuracy, illustrating the trade-off between model complexity and resource demands.

Advanced preprocessing methods significantly impact detection accuracy. 
For example, \cite{Zo8} uses the Frangi filter to suppress vessel-like structures, achieving 94\% sensitivity at the cost of high false positives (15.1/scan), whereas \cite{Z3} employs Adaptive Bilateral Filter (ABF) and ABC segmentation for precise lung region extraction, attaining 98.42\% accuracy. 
Conversely, \cite{Zo6} relies on K-means clustering and geometric mean filters, demonstrating that simpler techniques suffice for segmentation but limit nodule characterization precision.
Large-scale datasets like NLST in \cite{S8} (AUC: 94.4\%–95.5\%) and multi-institutional trials in \cite{Zo3} (96,559 participants) enhance generalizability, whereas smaller datasets in \cite{Zo8} and \cite{S11} risk overfitting despite high sensitivity (80.06\%–94\%). 
Notably, \cite{S11}’s two-step approach (geometric candidate generation + 3D CNN) mitigates this by combining prior knowledge with learned features, outperforming DBNs and SDAEs.

Methods prioritizing sensitivity, such as \cite{Zo8} (94\% sensitivity), incur higher false positives, while those emphasizing specificity, like \cite{S5} (91\% specificity), risk missing subtle nodules. 
\cite{Z5}’s DITNN classifier achieves 98.42\% accuracy with minimal error (0.038), illustrating how hybrid segmentation (IPCT) and noise reduction can harmonize both metrics. 
LDCT-focused studies like \cite{Zo3} demonstrate reduced lung cancer mortality but highlight unresolved challenges in overall mortality reduction. 
Meanwhile, \cite{S8}’s risk-prediction model leverages longitudinal LDCT data, underscoring the value of temporal analysis in early detection.
CT-based approaches excel in spatial accuracy but grapple with computational costs, false positives, and dataset biases. 
While 3D CNNs and hybrid models (e.g., \cite{Z4}, \cite{Z6}) dominate in performance, simpler pipelines (e.g., \cite{Zo7}) offer pragmatic alternatives for resource-constrained settings. 
Table II synthesizes these limitations, advocating for optimized preprocessing, federated learning, and explainability to bridge clinical adoption gaps.

\begin{table}

    \caption{CT Scans}
    \label{tab:my_label2}

    \centering
    \setlength{\tabcolsep}{1mm}
    \newcolumntype{M}[1]{>{\centering\arraybackslash}m{#1}}
    \begin{tabular}{|M{0.5cm}|M{2.75cm}|M{4.75cm}|}
    \hline 
         Ref. &  Methodology &  Disadvantages\\ \hline
         \cite{Z4}&  3D CNN and U-Net.&  High false positive rates. Relies on large labeled datasets for generalizability.\\ \hline
         \cite{S5}&  3D CNNs for Nodule Detection. &  Nodule Type Exclusion: Excludes certain types (e.g., GGO) that limit clinical applicability. 
         Small Datasets: Limited representation of nodule variability.\\ \hline 
         \cite{S10}&  Hybrid models for various DL models  &  
         Limited Depth: This may restrict feature learning capabilities, affecting diagnosis accuracy.
         Training Efficiency: Longer training times due to model complexity.\\ \hline
         \cite{Z2}&  Combining CAD with deep learning model and SVM.& 
         False Positives: High rates leading to unnecessary invasive procedures. 
         Radiation Exposure: Risk associated with multiple scans.\\ \hline 
         \cite{S8}& 3D CNN architecture & 
         False Positives: High rates leading to unnecessary invasive procedures. 
        Radiation Exposure: Risk associated with multiple scans.\\ \hline
         \cite{Zo6}&  ML-based approach like ANN, KNN, and RF &  False Negatives: Challenges in identifying smaller nodules or early-stage cancer. 
    Computational Complexity: Certain models require high computational resources.\\ \hline 
         \cite{Z6}&  Improved deep neural networks (IDNN) along with an ensemble classifier & Requires large datasets, risking overfitting, and high computational demands.\\ \hline  
         \cite{Zo3}&  Segmentation of the images was achieved through the K-means algorithm, and ML classifiers like ANN, KNN, and RF & False Negatives: Challenges in identifying smaller nodules or early-stage cancer. 
    Computational Complexity: Certain models require high computational resources.\\ \hline 
         \cite{Zo7}&  Deep neural network approach (MobileNet) &  
    Sensitivity Issues: Risk of misclassifying nodules, especially in early-stage cases.\\ \hline
         \cite{Z9}&  CNNs for feature selection and classification &  
         Small dataset of only 100 images.
         High computational complexity for real-time applications.\\ \hline
         \cite{Zo8} &  Multi-group patch-based CNN &  
         False Positives: Higher rates of false positives due to vessel-like structures in CT images.
         Patch Selection Challenges: Selecting meaningful patches can be computationally expensive.\\ \hline
         \cite{S11}&  3D CAD system using 3D CNNs &  Cross-Validation Limitations: Testing solely on training data can lead to overfitting.
         External Validation Needs: Lack of testing on independent datasets affects reliability.\\ \hline
         \cite{Z3}&  Conventional CT Scans combined with image processing techniques. & 
         Limited generalizations due to single hospital data.
         Requires significant processing power.\\ \hline         \cite{Z5}&  3D CNNs  &  
         Excludes certain nodule types.
         Small datasets limit variability.\\ \hline
    \end{tabular}

\end{table}

\subsection{Others}

\subsubsection{Whole Slide Images (WSI)}

WSI-based approaches leverage histopathology images to classify lung cancer subtypes and predict molecular markers, though they vary in supervision levels and clinical integration. 
The Inception v3 model in \cite{S3} achieves high diagnostic precision (AUC: 0.97) for NSCLC subtyping and gene mutation prediction using fully annotated TCGA datasets, but its dependency on exhaustive labeling limits scalability. 
In contrast, \cite{Zo9} adopts weakly supervised learning with patch-based FCNs and RF classifiers, attaining 97.3\% accuracy on 939 WSIs with minimal annotations. 
This method underperforms on public datasets (85.6\% AUC), reflecting a trade-off between annotation effort and generalizability. 
Meanwhile, \cite{Zo2} reviews broader AI applications in pathology, emphasizing CNNs for automated feature extraction but highlighting unresolved challenges, such as pathologist skepticism and workflow incompatibility.
WSI methods excel in molecular and histological granularity but face barriers in annotation standardization and clinical trust (Table III).

\subsubsection{PET Scan}

PET/CT frameworks prioritize multimodal integration and staging accuracy but grapple with technical artifacts and validation gaps. 
The hybrid CNN model in \cite{S9} combines PET and CT data to classify FDG uptake, achieving near-perfect AUC (0.99) for lung cancer and lymphoma, though its narrow focus limits broader clinical utility. 
Conversely, \cite{S19} proposes a cloud-based framework (Cloud-LTDSC) for end-to-end tumor detection, segmentation, and 9-stage classification, achieving 97\% accuracy on 94 NSCLC patients. 
While scalable, its small cohort risks overfitting. 
Clinical reviews contextualize PET/CT’s strengths, \cite{hochhegger2015pet} and \cite{Zo22} validate its superiority in nodal staging and lesion differentiation (e.g., T3/T4 staging) but note persistent false positives, necessitating invasive confirmations. 
\cite{zo23} further critiques technical limitations, such as respiration artifacts and contrast variability, advocating for protocol standardization. 
Though PET/CT excels in metabolic and anatomical correlation, its clinical adoption hinges on artifact mitigation and larger validation cohorts (Table III).

\begin{table}
    \caption{Others}
    \label{tab:my_label3}
    \centering
    \setlength{\tabcolsep}{1mm}
    \newcolumntype{M}[1]{>{\centering\arraybackslash}m{#1}}
    \begin{tabular}{|M{0.5cm}|M{2.75cm}|M{4.75cm}|} \hline 
         Ref. & Methodology &  Disadvantages\\ \hline 
         \cite{S3}&  PET/CT Scans. & Variability: Inconsistencies due to differing imaging protocols across institutions. 
         False Positives: Higher rates in non-malignant cases.\\ \hline 
         \cite{Zo9}& Weakly Supervised DL for Whole Slide Lung Cancer Image Analysis. FCNs for image segmentation and classification & Annotation Scarcity: Limited availability of labeled data leads to potential inaccuracies. 
         High Computational Costs: Processing whole slide images can be resource-intensive.\\ \hline
         \cite{Zo2} &  CNN for Pathology Image Analysis &  Large Data Requirement: CNNs require a large amount of annotated data for accurateness.
    Sensitivity to Variations: The model’s performance may degrade with variations in image quality.\\ 
         \hline
         \cite{S9}&  Automated Detection with PET/CT. CNNs for detecting and classifying FDG uptake in lung lesions &  Single Institution Data: Limited generalizations and potential biases. 
         False Positives: Frequent in lymphoma, complicating diagnosis.\\ \hline 
         \cite{hochhegger2015pet} & A standard combine PET and CT technologies & Radiation Exposure: expose patients to high radiation exposures.
         False Positives/Negatives: On PET scans, inflammatory diseases can occasionally appear as cancer, while, tiny lesions could go undetected.
         Cost and Availability: PET/CT scans can be costly, and not available.\\ \hline
         \cite{S19} & A cloud-based system and a multilayer convolutional neural network (M-CNN). & Concerns about the cloud privacy, the internet connection, and the computational cost.\\ \hline
         \cite{Zo22} & PET/CT. &
         False positives: that lead to unnecessary biopsies.
         High cost: PET/CT scans are expensive.
        Radiation exposure: higher radiation levels than individual PET or CT scans.\\ \hline
         \cite{zo23} & The integration of PET and CT for staging non-small cell lung cancer (NSCLC). & Concerns about the system availability, image interpretation complexity, and overdiagnosis.\\ \hline

    \end{tabular}

\end{table}

\section{Discussion and Open Problems}
\label{discussion}

High-resolution CT scans have shown strong potential for early detection of lung nodules, with studies reporting classification accuracies of up to 91.13\% \cite{Z4, S5} and even reaching 96\% in certain cases \cite{S18}. 
PET/CT imaging provides excellent specificity in identifying malignancies \cite{S9}, although its high resource demands and limited accessibility remain concerns. 
In contrast, X-ray imaging is widely available but suffers from sensitivity issues and false negatives caused by the poor detection of small nodules and interference from overlapping anatomical structures \cite{S4, Zo1}. 
WSI is also challenged by variability in imaging protocols and the occurrence of false positives in non-malignant cases \cite{S3, Zo9}.

Despite these promising findings, several limitations and open research problems persist. 
A primary challenge is the integration of imaging-based diagnostic systems into clinical workflows. 
Furthermore, the effective integration of multimodal data—combining imaging with genomic or clinical information—remains in its infancy. 
CT imaging face issues such as false positives, radiation exposure, and difficulties in detecting smaller nodules, as noted in multiple studies \cite{Z2, Z6, S8, Zo3, Zo6, Zo7}. 
Similarly, PET scans are limited by generalizability concerns due to reliance on single-institution data, and occasional false positives, especially in lymphoma detection \cite{S9}.

Enhancing image preprocessing and standardizing imaging protocols may improve diagnostic precision across modalities. 
For X-rays, the application of advanced image enhancement techniques could help reduce false negatives and misclassifications. 
In the case of CT scans, the adoption of low-dose imaging protocols and optimized preprocessing techniques may lower radiation exposure and decrease false-positive rates, while diversifying datasets and conducting extensive cross-validation could enhance the generalizability of PET imaging. 
For WSI, standardization of imaging protocols and the development of more efficient processing methods are recommended to overcome issues related to annotation scarcity and protocol variability. 
Overall, continued research focused on refining these imaging modalities and their integration into clinical practice is essential for advancing lung cancer detection.

This review of imaging modalities for lung cancer detection underscores the importance of early and accurate diagnosis. While 3D CNN architectures with CT scans show superior performance, critical challenges remain. Future research should include Creating standardized, diverse datasets to improve model generalizability across populations, and developing hybrid models to reduce false positives while maintaining sensitivity, particularly for CT-based approaches, and Optimizing computational efficiency for clinical implementation. Integrating multimodal imaging (CT with PET or WSI) and incorporating genetic data alongside imaging features presents promising directions. Additionally, explainable AI will be essential for clinical adoption. 

\section{Acknowledgment} 
The Security and Privacy of Emerging Network (SPEN) research lab at the University of Southern Mississippi (USM), USA, made this work possible. 
The statements made herein are solely the responsibility of the authors.

\bibliographystyle{ieeetr}

\begin{thebibliography}{10}

\bibitem{Zo9}
X.~Wang, H.~Chen, C.~Gan, H.~Lin, Q.~Dou, E.~Tsougenis, Q.~Huang, M.~Cai, and P.-A. Heng, ``Weakly supervised deep learning for whole slide lung cancer image analysis,'' {\em IEEE Trans. Cybern.}, vol.~50, pp.~3950--3962, Sept. 2020.

\bibitem{S10}
W.~Sun, B.~Zheng, and W.~Qian, ``Automatic feature learning using multichannel roi based on deep structured algorithms for computerized lung cancer diagnosis,'' {\em Computers in Biology and Medicine}, vol.~89, pp.~530--539, 2017.

\bibitem{nn1}
B.~Hamoui, A.~Alashaikh, A.~Sherif, E.~Alanazi, M.~Nabil, and W.~Alsmary, ``Google searches and covid-19 cases in saudi arabia: A correlation study,'' in {\em 2021 3rd IEEE Middle East and North Africa COMMunications Conference (MENACOMM)}, pp.~104--108, 2021.

\bibitem{nn2}
C.~Bourn, J.~Heirendt, K.~Ben-Chiobi, A.~Haastrup, A.~Sherif, and M.~Elsersy, ``Privacy-preserving data sharing scheme for e-health systems,'' in {\em 2023 9th International Conference on Information Technology Trends (ITT)}, pp.~20--25, 2023.

\bibitem{nn3}
J.~Romeo, M.~Abbass, A.~Sherif, M.~M.~R. Khan~Mamun, M.~Elsersy, and K.~Khalil, ``Privacy-preserving machine learning for e-health applications: A survey,'' in {\em 2024 IEEE 3rd International Conference on Computing and Machine Intelligence (ICMI)}, pp.~1--6, 2024.

\bibitem{nn4}
M.~Watkins, C.~Dorsey, D.~Rennier, T.~Polley, A.~Sherif, and M.~Elsersy, ``Privacy-preserving data aggregation scheme for e-health,'' in {\em Proceedings of the 2nd International Conference on Emerging Technologies and Intelligent Systems} (M.~A. Al-Sharafi, M.~Al-Emran, M.~N. Al-Kabi, and K.~Shaalan, eds.), (Cham), pp.~638--646, Springer International Publishing, 2023.

\bibitem{nn5}
M.~Elsersy, A.~Sherif, A.~A.-A. Imam, M.~M.~R. Khan~Mamun, K.~Khalil, and M.~Haitham, ``Federated learning model for early detection of dementia using blood biosamples,'' in {\em 2023 IEEE International Conference on Artificial Intelligence, Blockchain, and Internet of Things (AIBThings)}, pp.~1--5, 2023.

\bibitem{Z10}
G.~Chassagnon, C.~De~Margerie-Mellon, M.~Vakalopoulou, R.~Marini, T.-N. Hoang-Thi, M.-P. Revel, and P.~Soyer, ``Artificial intelligence in lung cancer: current applications and perspectives,'' {\em Jpn. J. Radiol.}, vol.~41, pp.~235--244, Mar. 2023.

\bibitem{Z11}
M.~Cellina, L.~M. Cacioppa, M.~C{\`e}, V.~Chiarpenello, M.~Costa, Z.~Vincenzo, D.~Pais, M.~V. Bausano, N.~Rossini, A.~Bruno, and C.~Floridi, ``Artificial intelligence in lung cancer screening: The future is now,'' {\em Cancers (Basel)}, vol.~15, Aug. 2023.

\bibitem{Z12}
H.~T. Gayap and M.~A. Akhloufi, ``Deep machine learning for medical diagnosis, application to lung cancer detection: A review,'' {\em BioMedInformatics}, vol.~4, no.~1, pp.~236--284, 2024.

\bibitem{S4}
Z.~Peng, X.~Xinnan, W.~Hongwei, F.~Yuanli, F.~Haozhe, Z.~Jianwei, Y.~Shoukun, H.~Yuxuan, S.~Yiwen, L.~Jiaxiang, and L.~Xinguo, ``Computer-aided lung cancer diagnosis approaches based on deep learning,'' {\em Journal of Computer-Aided Design \& Computer Graphics}, vol.~30, no.~1, pp.~90--99, 2018.

\bibitem{li2020exploring}
K.~Li, Y.~Chen, R.~Sun, B.~Yu, G.~Li, and X.~Jiang, ``Exploring potential of different x-ray imaging methods for early-stage lung cancer detection,'' {\em Radiation Detection Technology and Methods}, vol.~4, pp.~213--221, 2020.

\bibitem{S16}
I.~S. Abed, ``Lung cancer detection from x-ray images by combined backpropagation neural network and pca,'' {\em Engineering and Technology Journal}, vol.~37, no.~5A, pp.~166--171, 2019.

\bibitem{S14}
M.~F. Syahputra, R.~F. Rahmat, and R.~Rambe, ``Identification of lung cancer on chest x-ray (cxr) medical images using the probabilistic neural network method,'' {\em Journal of Physics: Conference Series}, vol.~1898, p.~012023, jun 2021.

\bibitem{S15}
K.~Rajagopalan and S.~Babu, ``The detection of lung cancer using massive artificial neural network based on soft tissue technique,'' {\em BMC Medical Informatics and Decision Making}, vol.~20, p.~282, Oct 2020.

\bibitem{S13}
H.~Malik, T.~Anees, M.~Din, and A.~Naeem, ``Cdc{\_}net: multi-classification convolutional neural network model for detection of covid-19, pneumothorax, pneumonia, lung cancer, and tuberculosis using chest x-rays,'' {\em Multimedia Tools and Applications}, vol.~82, pp.~13855--13880, Apr 2023.

\bibitem{Zo1}
D.~M. Ibrahim, N.~M. Elshennawy, and A.~M. Sarhan, ``Deep-chest: Multi-classification deep learning model for diagnosing covid-19, pneumonia, and lung cancer chest diseases,'' {\em Computers in Biology and Medicine}, vol.~132, p.~104348, 2021.

\bibitem{S18}
V.~Sreeprada and K.~Vedavathi, ``Lung cancer detection from x-ray images using hybrid deep learning technique,'' {\em Procedia Computer Science}, vol.~230, pp.~467--474, 2023.

\bibitem{S17}
S.~Bharati, P.~Podder, and M.~R.~H. Mondal, ``Hybrid deep learning for detecting lung diseases from x-ray images,'' {\em Informatics in Medicine Unlocked}, vol.~20, p.~100391, 2020.

\bibitem{S12}
J.~Juan, E.~Mons{\'o}, C.~Lozano, M.~Cuf{\'i}, P.~Sub{\'i}as-Beltr{\'a}n, L.~Ruiz-Dern, X.~Rafael-Palou, M.~Andreu, E.~Casta{\~{n}}er, X.~Gallardo, A.~Ullastres, C.~Sans, M.~Luj{\`a}n, C.~Rubi{\'e}s, and V.~Ribas-Ripoll, ``Computer-assisted diagnosis for an early identification of lung cancer in chest x rays,'' {\em Scientific Reports}, vol.~13, p.~7720, May 2023.

\bibitem{Z4}
W.~Alakwaa, M.~Nassef, and A.~Badr, ``Lung cancer detection and classification with 3d convolutional neural network (3d-cnn),'' {\em International Journal of Advanced Computer Science and Applications}, vol.~8, no.~8, 2017.

\bibitem{S5}
N.~Nasrullah, J.~Sang, M.~S. Alam, M.~Mateen, B.~Cai, and H.~Hu, ``Automated lung nodule detection and classification using deep learning combined with multiple strategies,'' {\em Sensors}, vol.~19, no.~17, 2019.

\bibitem{Z6}
P.~M. Shakeel, M.~A. Burhanuddin, and M.~I. Desa, ``Automatic lung cancer detection from ct image using improved deep neural network and ensemble classifier,'' {\em Neural Computing and Applications}, vol.~34, pp.~9579--9592, Jun 2022.

\bibitem{Z9}
M.~Toƒüa√ßar, B.~Ergen, and Z.~C√∂mert, ``Detection of lung cancer on chest ct images using minimum redundancy maximum relevance feature selection method with convolutional neural networks,'' {\em Biocybernetics and Biomedical Engineering}, vol.~40, no.~1, pp.~23--39, 2020.

\bibitem{Zo7}
P.~Shill and Z.~Homayra, ``A new method for lung nodule detection using deep neural networks for ct images,'' pp.~1--6, 02 2019.

\bibitem{Zo8}
H.~Jiang, H.~Ma, W.~Qian, M.~Gao, and Y.~Li, ``An automatic detection system of lung nodule based on multigroup patch-based deep learning network,'' {\em IEEE J. Biomed. Health Inform.}, vol.~22, pp.~1227--1237, July 2018.

\bibitem{Z3}
A.~Asuntha and A.~Srinivasan, ``Deep learning for lung cancer detection and classification,'' {\em Multimedia Tools and Applications}, vol.~79, pp.~7731--7762, Mar 2020.

\bibitem{Zo6}
S.~Nageswaran, G.~Arunkumar, A.~K. Bisht, S.~Mewada, J.~N. V. R.~S. Kumar, M.~Jawarneh, and E.~Asenso, ``Lung cancer classification and prediction using machine learning and image processing,'' {\em Biomed Res. Int.}, vol.~2022, p.~1755460, Aug. 2022.

\bibitem{S8}
D.~Ardila, A.~P. Kiraly, S.~Bharadwaj, B.~Choi, J.~J. Reicher, L.~Peng, D.~Tse, M.~Etemadi, W.~Ye, G.~Corrado, D.~P. Naidich, and S.~Shetty, ``End-to-end lung cancer screening with three-dimensional deep learning on low-dose chest computed tomography,'' {\em Nature Medicine}, vol.~25, pp.~954--961, Jun 2019.

\bibitem{Zo3}
R.~M. Hoffman, R.~P. Atallah, R.~D. Struble, and R.~G. Badgett, ``Lung cancer screening with low-dose {CT}: A meta-analysis,'' {\em J. Gen. Intern. Med.}, vol.~35, pp.~3015--3025, Oct. 2020.

\bibitem{S11}
X.~Huang, J.~Shan, and V.~Vaidya, ``Lung nodule detection in ct using 3d convolutional neural networks,'' in {\em 2017 IEEE 14th International Symposium on Biomedical Imaging (ISBI 2017)}, pp.~379--383, 2017.

\bibitem{Z5}
P.~M. Shakeel, M.~Burhanuddin, and M.~I. Desa {\em Measurement}, vol.~145, pp.~702--712, 2019.

\bibitem{Z2}
S.~Makaju, P.~Prasad, A.~Alsadoon, A.~Singh, and A.~Elchouemi, ``Lung cancer detection using ct scan images,'' {\em Procedia Computer Science}, vol.~125, pp.~107--114, 2018.
\newblock The 6th International Conference on Smart Computing and Communications.

\bibitem{S3}
N.~Coudray, P.~S. Ocampo, T.~Sakellaropoulos, N.~Narula, M.~Snuderl, D.~Feny{\"o}, A.~L. Moreira, N.~Razavian, and A.~Tsirigos, ``Classification and mutation prediction from non--small cell lung cancer histopathology images using deep learning,'' {\em Nature Medicine}, vol.~24, pp.~1559--1567, Oct 2018.

\bibitem{Zo2}
S.~Wang, D.~M. Yang, R.~Rong, X.~Zhan, J.~Fujimoto, H.~Liu, J.~Minna, I.~I. Wistuba, Y.~Xie, and G.~Xiao, ``Artificial intelligence in lung cancer pathology image analysis,'' {\em Cancers (Basel)}, vol.~11, p.~1673, Oct. 2019.

\bibitem{S9}
L.~Sibille, R.~Seifert, N.~Avramovic, T.~Vehren, B.~Spottiswoode, S.~Zuehlsdorff, and M.~Sch\"{a}fers, ``18f-fdg pet/ct uptake classification in lymphoma and lung cancer by using deep convolutional neural networks,'' {\em Radiology}, vol.~294, no.~2, pp.~445--452, 2020.
\newblock PMID: 31821122.

\bibitem{S19}
G.~Kasinathan and S.~Jayakumar, ``Cloud-based lung tumor detection and stage classification using deep learning techniques,'' {\em Biomed research international}, vol.~2022, no.~1, p.~4185835, 2022.

\bibitem{hochhegger2015pet}
B.~Hochhegger, G.~R.~T. Alves, K.~L. Irion, C.~C. Fritscher, L.~G. Fritscher, N.~H. Concatto, and E.~Marchiori, ``Pet/ct imaging in lung cancer: indications and findings,'' {\em Jornal Brasileiro de Pneumologia}, vol.~41, no.~3, pp.~264--274, 2015.

\bibitem{Zo22}
A.~F. E. A. M. D. A. F. C. C.~L. Cristina~Gámez, Rafael~Rosell, ``Pet/ct fusion scan in lung cancer: Current recommendations and innovations,'' {\em Journal of Thoracic Oncology}, vol.~11, no.~1, pp.~6--10, 2016.

\bibitem{zo23}
J.~C. J. A.~V. W~De~Wever, S~Stroobants, ``Integrated pet/ct in the staging of nonsmall cell lung cancer: technical aspects and clinical integration,'' {\em European Respiratory Journal}, vol.~33, no.~1, pp.~201--212, 2009.

\end{thebibliography}

\end{document}